\documentclass[12pt,preprint]{aastex}

\newcommand{\Dnu}[1]{\Delta \nu_{#1}}
\newcommand{\dnu}[1]{\delta \nu_{#1}}
\newcommand{\acena}{\mbox{$\alpha$~Cen~A}}
\newcommand{\acenb}{\mbox{$\alpha$~Cen~B}}
\newcommand{\acen}{\mbox{$\alpha$~Cen}}

\newcommand{\cms}{\mbox{cm\,s$^{-1}$}}

\newcommand{\muHz}{\mbox{$\mu$Hz}}
\newcommand{\note}[1]{{\bf [#1]}}
\renewcommand{\note}[1]{\relax}
\let\epsilon\varepsilon

\slugcomment{Accepted by ApJ}

\shorttitle{Oscillations in $\alpha$ Cen A}
\shortauthors{Bedding et al.}

\begin{document}

\title{Oscillation frequencies and mode lifetimes in $\alpha$~Centauri~A}

\author{
Timothy R. Bedding,\altaffilmark{1}
Hans Kjeldsen,\altaffilmark{2}
R. Paul Butler,\altaffilmark{3}
Chris McCarthy,\altaffilmark{3}
Geoffrey W. Marcy,\altaffilmark{4}
Simon J. O'Toole,\altaffilmark{1,5}
Christopher G. Tinney,\altaffilmark{6}
Jason T. Wright\altaffilmark{4}
}

\altaffiltext{1}{School of Physics A28, University of Sydney, NSW 2006,
Australia; bedding@physics.usyd.edu.au}

\altaffiltext{2}{Theoretical Astrophysics Center, University of Aarhus,
DK-8000 Aarhus C, Denmark; hans@phys.au.dk}

\altaffiltext{3}{Carnegie Institution of Washington,
Department of Terrestrial Magnetism, 5241 Broad Branch Road NW, Washington,
DC 20015-1305; paul@dtm.ciw.edu, chris@dtm.ciw.edu}

\altaffiltext{4}{Department of Astronomy, University of California,
Berkeley, CA 94720; and Department of Physics and Astronomy, San Francisco,
CA 94132; gmarcy@astron.berkeley.edu, jtwright@astron.berkeley.edu}

\altaffiltext{5}{Universit\"at Erlangen-N\"urnberg, Sternwartstr. 7,
D-96049 Bamberg, Germany; otoole@sternwarte.uni-erlangen.de}

\altaffiltext{6}{Anglo-Australian Observatory, P.O.\,Box 296, Epping, NSW
1710, Australia; cgt@aaoepp.aao.gov.au}

\begin{abstract} 
We analyse our recently-published velocity measurements of $\alpha$~Cen~A
(Butler et al.\ 2004).  After adjusting the weights on a night-by-night
basis in order to optimize the window function to minimize sidelobes, we
extract 42 oscillation frequencies with $l=0$--3 and measure the large and
small frequency separations.  We give fitted relations to these frequencies
that can be compared with theoretical models and conclude that the
observed scatter about these fits is due to the finite lifetimes of the
oscillation modes.  We estimate the mode lifetimes to be 1--2\,d,
substantially shorter than in the Sun.
\end{abstract}

\keywords{stars: individual ($\alpha$~Cen~A) --- stars: oscillations---
techniques: radial velocities\vspace*{13ex}}

\section{Introduction}

Its proximity and its similarity to the Sun have made $\alpha$~Cen~A a
prime target for asteroseismology.  The first clear detection of p-mode
oscillations was made by \citet{B+C2001,B+C2002} using velocity
measurements with the CORALIE spectrograph over 13 nights.  These
results confirmed the earlier but less secure detection by \citet{S+B2000},
who used photometry from the {\em WIRE\/} satellite.  Based on the CORALIE
data, \citet{B+C2002} published frequencies for 28 modes with $l = 0$--$2$
and $n=15$--$25$ and these have been compared with theoretical models by
\citet{TPM2002} and \citet{TSN2003}.  Meanwhile, oscillations in
\acenb{} have been reported by \citet{C+B2003}, adding to the
growing list of stars in which solar-like oscillations have been detected
-- see \citet{B+K2003} for a recent review.

We recently reported observations of \acena{} made over 5 nights in
both Chile with UVES\footnote{Based on observations collected at the
European Southern Observatory, Paranal, Chile (ESO Programme 67.D-0133)}
and in Australia with UCLES \citep[][hereafter Paper~I]{BBK2004}.  These
observations covered a shorter time span than those by \citeauthor{B+C2001}
but had substantially better velocity precision and the benefit of two-site
coverage.

In Paper~I we described the method used to process the velocity
measurements.  In brief, we first removed slow drifts and sudden jumps of a
few metres per second, presumably instrumental, from each time series.  We
then used the measurement uncertainties as weights in calculating the power
spectrum (according to $w_i = 1/\sigma_i^2$), but we found it necessary to
modify some of the weights to account for a small fraction of bad data
points.  In the end, we reached a noise floor of 2.0\,\cms{} in the
combined amplitude spectrum.  In this paper, we extract oscillation
frequencies, measure the large and small frequency separations and also
estimate the mode lifetimes.

\section{Optimizing the window function}

The observing window contains gaps due to day time and bad weather, which
means that each peak in the power spectrum is accompanied by sidelobes.
This can be quantified by calculating the spectral window, which is the
power spectrum of the observing window and is shown in Fig.~6 of Paper~I.
The two-site coverage of our observations means that the gaps are
relatively small in the latter part of the run, when the weather was best.
However, as discussed in Paper~I, the UVES data have much higher weight
than the UCLES data, due to their higher precision.  This means that the
spectral window, which takes the weights into account, has quite strong
sidelobes (24\% in power).  As is well known, these sidelobes complicate
the oscillation spectrum, especially for weaker modes, for which the
interaction between noise and signal can cause sidelobes to become higher
than the central peak and lead to mis-identifications.


To address this problem, we have adjusted the weights on a night-by-night
basis in order to optimize the window function.  To be specific, we
allocated separate adjustment factors to each of the three UVES nights and
the five UCLES nights.  The weights on each night were multiplied by these
factors and the spectral window was calculated, and this process was
iterated to minimize the height of the sidelobes.  The final values for
these weight multipliers were 0.88, 1.00 and 1.12 for the three UVES nights
and 0.11, 0.42, 3.25, 8.19 and 7.62 for the five UCLES nights.  We treated
separately the short period of overlap (see Paper~I), and the weight
multiplier for that period turned out to be~1.39.  

\begin{figure*}
\epsscale{0.9}
\hspace*{-2cm}
\plotone{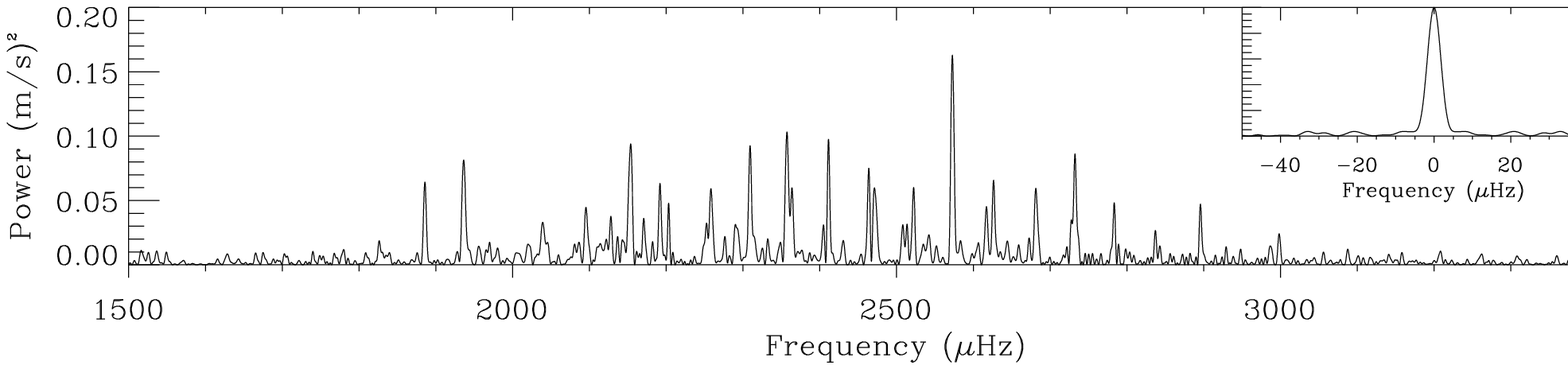}
\caption[]{\label{fig.power} Power spectrum of the combined \acena{}
velocity time series, with weights for each night adjusted to optimize the
window function.  The inset shows the spectral window (power in arbitrary
units), with the frequency scale expanded by a factor of~5. \note{Note to
editor: please typeset double-column}}
\end{figure*}

This optimization process has significantly increased the weight given to
the last three UCLES nights, which makes sense because these fill in the
gaps in the UVES data.  The power spectrum based on these adjustment
factors is shown in Fig.~\ref{fig.power}.  The resulting spectral window is
shown in the inset, and the highest sidelobes are only 3.6\% in power.
The optimization process has slightly degraded the frequency
resolution, with the FWHM of the spectral window increasing from
3.85\,\muHz{} to 4.13\,\muHz.  However, the main trade-off is an increase
in the noise floor in the amplitude spectrum at high frequencies, from 2.0
to 2.9\,\cms.  Nevertheless, we have found that the improvement in the
spectral window makes the oscillation spectrum much clearer, and this
compensates for the reduction in signal-to-noise.  We note that we found
similar results to those reported here when we performed the analysis using
a compromise weighting scheme, in which we used the square roots of the
weight multipliers given above.

%

\section{Frequency analysis}	\label{sec.freq}

\begin{figure}
\epsscale{0.4}
\plotone{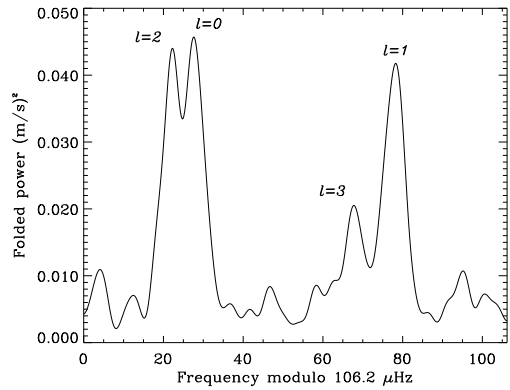}\\
\caption[]{\label{fig.folded} The power spectrum of \acena{} folded at the
large separation, showing peaks corresponding to $l=0$--3.}
\end{figure}

\begin{figure}
\epsscale{0.4}
\plotone{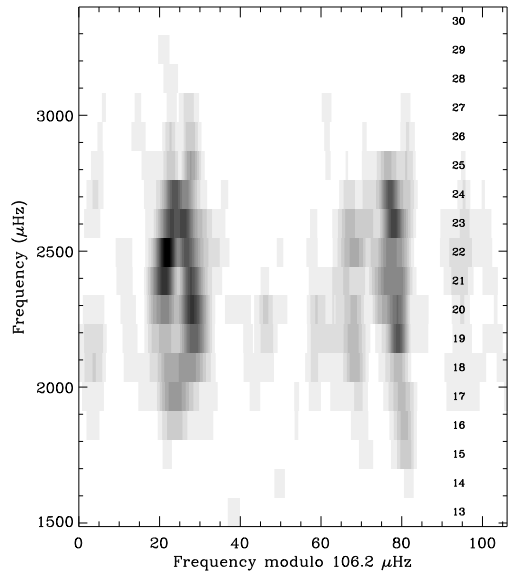}
\caption[]{\label{fig.smoothed.echelle} The power spectrum of \acena{} in
echelle format, and smoothed slightly in the vertical direction.  The
column of numbers on the right shows the $n$ value of each order.}
\end{figure}

The large separation of \acena{} was measured to be $\sim106\,$\muHz{} by
\citet{S+B2000} and \citet{B+C2001}.  In Fig.~\ref{fig.folded} we show the
central part of our power spectrum (2020--2970\,\muHz) folded with a
spacing of $\Delta\nu = 106.2$\,\muHz.  We can clearly see peaks
corresponding to modes with degree $l=0, 1, 2$ and~$3$.  In
Fig.~\ref{fig.smoothed.echelle} we show the power spectrum in echelle
format, where we have smoothed in the vertical direction to make the ridges
more visible.  Again, four ridges corresponding to $l=0$--$3$ are visible.

\begin{figure}
\epsscale{0.48}
\plotone{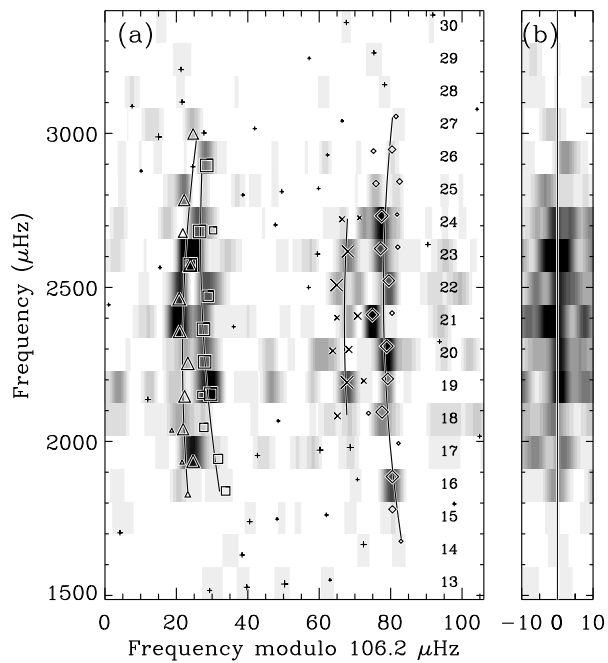}
\caption[]{\label{fig.echelle} (a)~The greyscale shows the power spectrum
in echelle format (without smoothing) and the symbols show the peaks
extracted by iterative sine-wave fitting: $l=0$ (squares), $l=1$
(diamonds), $l=2$ (triangles), $l=3$ (crosses) and other peaks (plus
signs).  The sizes of the symbols are proportional to the amplitudes of the
peaks.  The curves show fits to the frequencies, as described in the text,
and the numbers up the right side show the $n$ value of each order. (b)~The
power summed over the four ridges ($l=0$--3) after first subtracting from
each one the frequency predicted by
equations~(\ref{eq.nu0})--(\ref{eq.nu3}).  }
\end{figure}

The next step was to measure the frequencies of the strongest peaks in the
power spectrum, which we did in the standard way using iterative sine-wave
fitting.  These frequencies are given in Table~\ref{tab.freqs} and are
shown as symbols in Fig.~\ref{fig.echelle}(a), over-plotted on a grey scale
of the power spectrum itself (this time without any smoothing).  The sizes
of the symbols indicate the strengths of the peaks (the amplitudes are
given in Table~\ref{tab.amps}).  We have assigned $l$ values to the
strongest peaks, based on their location in the echelle diagram.  The four
curves are fits to the data and are described below.

\begin{figure}
\epsscale{0.8}
\plotone{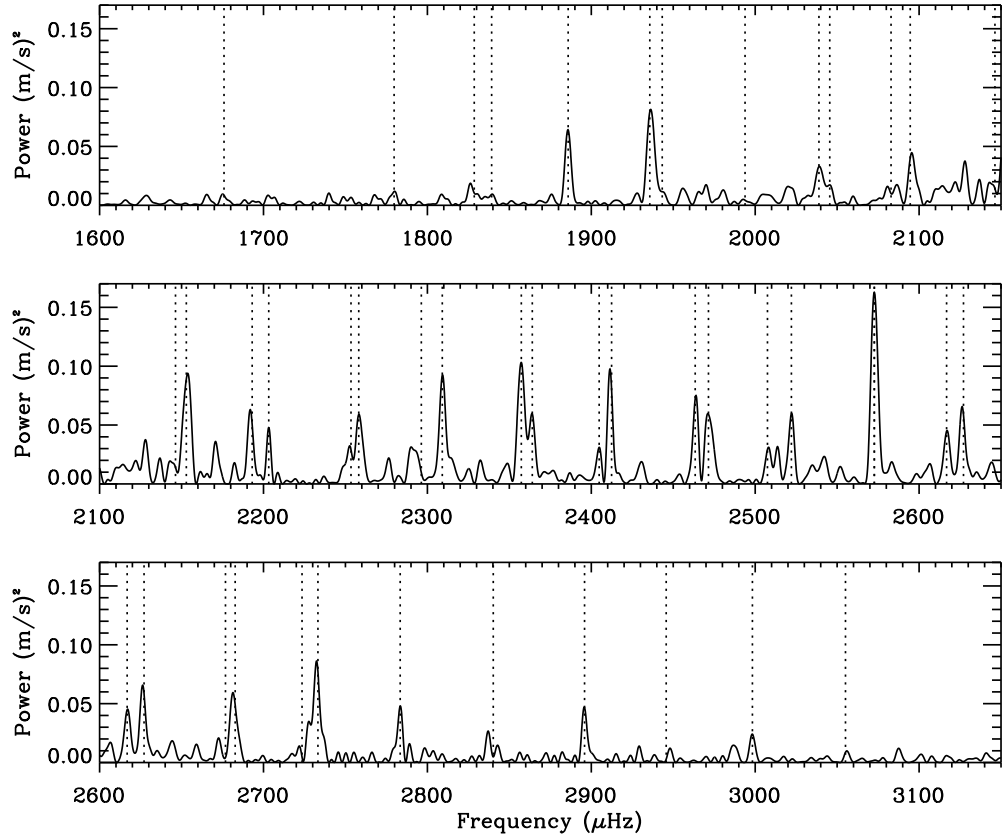}
\caption[]{\label{fig.power-freq} Power spectrum of \acena{} with mode
frequencies overlaid.  Same as Fig.~\ref{fig.power}, but on an expanded
scale and with the frequencies in Table~\ref{tab.freqs} marked by dotted
lines.}  \note{Note to editor: please typeset double-column}
\end{figure}

There is a significant scatter of the individual peaks about each ridge,
which we interpret as reflecting the finite lifetime of the oscillation
modes.  In several cases, the iterative sine-wave fitting has produced two
peaks that appear to represent a single oscillation mode, which is to be
expected if we have partially resolved the modes (see
Sec.~\ref{sec.lifetimes} for further discussion of the mode lifetimes).  In
these cases, both peaks are shown in Fig.~\ref{fig.echelle}(a) but in
Tables~\ref{tab.freqs} and~\ref{tab.amps} (and also in
Fig.~\ref{fig.power-freq}) they are combined into a single weighted mean.
It is also important to note that the separation between $l=0$ and $l=2$ is
not well resolved by our observations, and in one case (2572.7\,\muHz) it
seems that the two modes have combined to form a single peak.

To show the power spectrum more clearly, we have plotted it on an expanded
scale in Fig.~\ref{fig.power-freq} and indicated the extracted frequencies.
Note that interference between noise peaks, oscillation peaks and
their sidelobes has created apparent mismatches in a few places between the
peaks in the power spectrum and the dotted lines representing the extracted
frequencies.  Such interference is relatively minor given the low level of
both noise and sidelobes, but it is still significant (and it is, of
course, the reason we chose to extract frequencies using the iterative
fitting process).  Also, as mentioned above, the iterative fitting has
sometimes produced two peaks that have been averaged into a single value for
this figure.

Asteroseismology involves comparing observed frequencies with theoretical
models, and the values listed in Table~\ref{tab.freqs} could be used for
that purpose.  However, the scatter caused by the finite mode lifetimes
means that we would do better to fit to the ridges, in order to obtain more
accurate estimates for the eigenfrequencies of the star.  We now describe
the procedure that we used to perform this fit, which begins with an
examination of the small frequency separations.

\begin{figure}
\epsscale{0.48}
\plotone{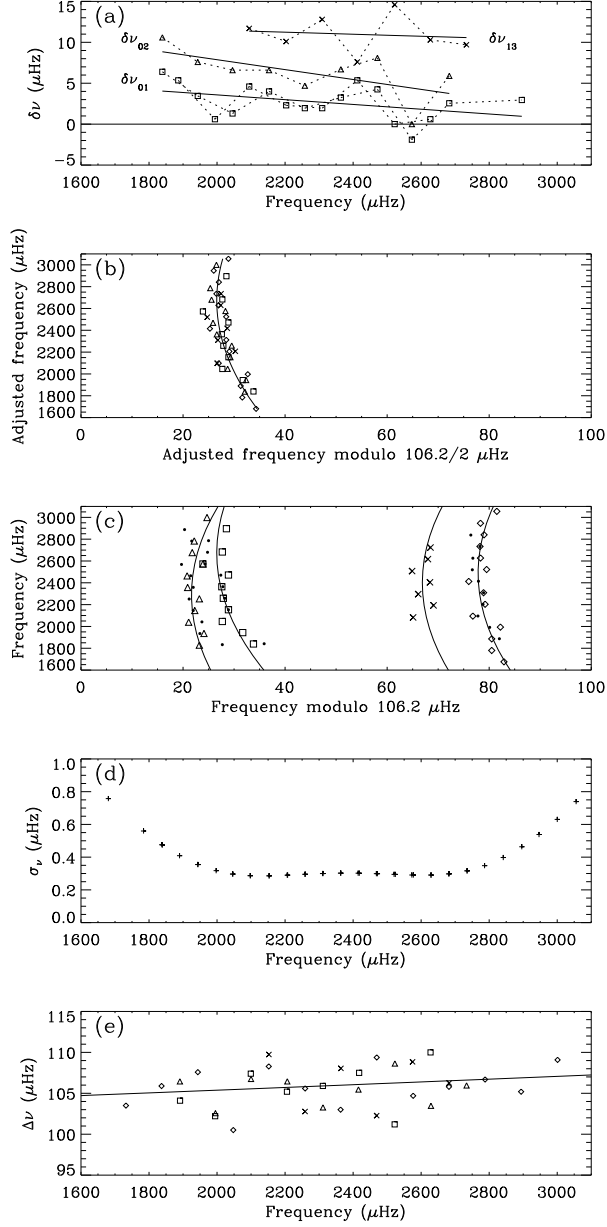}
\caption[]{\label{fig.fits} (a)~The three small separations as a function
of frequency.  Symbols connected by dotted lines show the measured values:
$\dnu{02}$ (triangles), $\dnu{13}$ (crosses) and $\dnu{01}$ (two sets of
squares).  The three solid lines are lines of best fit. (b)~The frequencies
after adjusting by subtracting off the small separations, to align them in
in a single ridge.  The curve is the parabola of best fit. (c)~The measured
frequencies (unadjusted) in echelle format.  The curves show the four
parabolas defined by equations~(\ref{eq.nu0})--(\ref{eq.nu3}), and the
small solid circles show the frequencies reported by \citet{B+C2002}.
(d)~Formal uncertainty in the fit as a function of frequency.  (e)~The
large separation as a function of frequency for the four values of~$l$,
where the line shows the separation for $l=1$ calculated from
equation~(\ref{eq.nu1}).  In panels (b), (c) and (e) the symbols indicate
$l$ values, with the same meaning as in Fig.~\ref{fig.echelle}.  }
\end{figure}

Figure~\ref{fig.fits}(a) shows our measured values for the three small
separations, which are defined according to the usual conventions
\citep[for a recent review, see][]{B+K2003}. Thus, $\dnu{02}$ is the
separation between adjacent peaks with $l=0$ and $l=2$ and $\dnu{13}$ is
the separation between $l=1$ and $l=3$.  The third small separation,
$\dnu{01}$, is the amount by which $l=1$ modes are offset from the midpoint
between the $l=0$ modes on either side.  Since one could equally well
define $\dnu{01}$ to be the offset of $l=0$ modes from the midpoint between
consecutive $l=1$ modes, we have shown both versions in
Fig.~\ref{fig.fits}(a).

The observations show that the small separations in \acen{} vary with
frequency, as was already reported for $\dnu{02}$ by \citet{B+C2002}.  We
have fitted straight lines to each of the three separations (fitting a
polynomial of higher degree is not justified by the data).  These lines are
shown in Fig.~\ref{fig.fits}(a) and the equations are as follows:
\begin{eqnarray}
 \dnu{02} & = & ~~5.46\,\muHz \;-\; 0.0060 \;(\nu - 2400\,\muHz) \\
 \dnu{13} & = &  10.99\,\muHz \;-\; 0.0012 \;(\nu - 2400\,\muHz) \\
 \dnu{01} & = & ~~2.41\,\muHz \;-\; 0.0029 \;(\nu - 2400\,\muHz)
\end{eqnarray}
Using these fits, we were then able to adjust the measured frequencies to
remove the effect of the small separations and so align them all in a
single ridge.  The result is shown in Fig.~\ref{fig.fits}(b).  Note that we
have taken the frequencies modulo $\Dnu{}/2$, so that the even ($l=0$, 2)
and odd ($l=1$, 3) ridges are superimposed.  To this single ridge we then
fitted a parabola (the highest degree polynomial that is justified by the
data), which is shown in the figure.

This completes our fit to the observed frequencies, which has involved nine
free parameters (three linear fits and one parabola).  We show this fit and
the observed frequencies in Fig.~\ref{fig.fits}(c), and also in
Fig.~\ref{fig.echelle}(a).  The equations for this fit are as follows:
\begin{eqnarray}
 \nu_{n,0} & = &    \left(2363.85 + 105.70 \;(n - 21) + 0.088 \;(n-21)^2\right)\,\muHz \label{eq.nu0}\\
 \nu_{n,1} & = &    \left(2414.35 + 106.10 \;(n - 21) + 0.088 \;(n-21)^2\right)\,\muHz \label{eq.nu1}\\
 \nu_{n,2} & = &    \left(2464.60 + 106.52 \;(n - 21) + 0.088 \;(n-21)^2\right)\,\muHz \label{eq.nu2}\\
 \nu_{n,3} & = &    \left(2509.70 + 106.41 \;(n - 21) + 0.088 \;(n-21)^2\right)\,\muHz.\label{eq.nu3} 
\end{eqnarray}
These relations, rather than the actual frequencies, should be used when
comparing with theoretical models because the measured frequencies are
significantly affected by the mode lifetime.  Note that these relations
should not be extrapolated to frequencies beyond those actually measured.
The formal uncertainties on the frequencies given by these relations are
shown in Fig.~\ref{fig.fits}(d), based on the uncertainties in the
parameters of the parabola that was fitted in Fig.~\ref{fig.fits}(b).
Over most of the range the uncertainty is 0.3--0.4\,\muHz, rising to about
0.8\,\muHz{} at the ends.

To examine whether the oscillations frequencies deviate systematically from
equations~(\ref{eq.nu0})--(\ref{eq.nu3}), we have used those equations to
``straighten'' the ridges.  Figure~\ref{fig.echelle}(b) shows the result,
where we have summed the power of the four ridges ($l=0$--3) after first
subtracting from each the frequency predicted by
equations~(\ref{eq.nu0})--(\ref{eq.nu3}).  Looking at the highest
frequencies (above 3000\,\muHz), there seems to be evidence for a departure
from the fit in the sense that the ridge of power curves to the left.  More
observations are needed to confirm this result.

Another useful parameter to compare with models is the large
separation,~$\Dnu{}$.  The measured values are shown in
Fig.~\ref{fig.fits}(e) and there is an upward trend with frequency that was
previously noted by \citet{B+C2002} and which is responsible for the
curvature in the echelle diagram.  The line in Fig.~\ref{fig.fits}(e) shows
the large separation for $l=1$ calculated from equation~(\ref{eq.nu1}).
Lines for other $l$ values (not shown) are almost indistinguishable.

The small solid circles in Fig.~\ref{fig.fits}(c) show the 28 frequencies
reported by \citet{B+C2002}.  Note these are all $l = 0$--$2$ because those
observations were single-site and aliases prevented the detection of $l=3$.
On the other hand, our observations were shorter, which made it more
difficult to separate $l=0$ from $l=2$.  Keeping in mind the scatter
induced by the finite mode lifetimes, the agreement between the two results
is excellent, especially for $l=1$.

\section{Mode lifetimes}        \label{sec.lifetimes}

As already mentioned, the observed frequencies show a significant scatter
about the fitted relations.  If the modes were coherent (i.e., pure
sinusoids with infinite lifetimes), we have shown using simulations that we
would be able to measure them with a typical accuracy of 0.4\,\muHz{} or
better.  The observed scatter is much greater than this, which we interpret
as reflecting the finite lifetimes of the oscillation modes.  As is well
established for the Sun, the power spectrum of a stochastically excited
oscillation that is observed for long enough will display a series of
closely spaced peaks under a Lorentzian profile, the width of which
indicates the mode lifetime \citep[e.g.,][]{T+F92}.  If, as in this case,
the observations are not long enough to resolve the Lorentzian profile then
the effect of the finite mode lifetime is to randomly shift each
oscillation peak from its true position by a small amount.  This is
responsible for the scatter observed in Fig.~\ref{fig.fits}.

Measuring this scatter gives us an opportunity to estimate the mode
lifetime in \acena.  
%
%
In Table~\ref{tab.lifetime} we give the scatter of the observed frequencies
about the fit for two frequency ranges (a finer subdivision in frequency is
not justified by the data).
We have converted these to mode lifetimes using simulations, as
follows.  We simulated a single oscillation mode, sampled with the same
window function and weights as the actual observations.  We used the method
described by \citet{SKB2004} to simulate an oscillation that was re-excited
continuously with random kicks and damped on a timescale that was an
adjustable parameter (the mode lifetime).  For a given mode lifetime, we
created 100 simulated time series that differed only in the random number
seed.  For each one we measured the position of the peak in the power
spectrum and hence determined the scatter of the ``observed'' frequency
about its true value.  This was repeated for different values of the mode
lifetime, allowing us to calibrate the relationship between the mode
lifetime and the scatter in the observed frequencies.  The inferred mode
lifetimes for \acena{} are given in Table~\ref{tab.lifetime}.

To check this technique, we also analysed segments of an 805-day series of
full-disk velocity observations of the Sun taken by the GOLF instrument on
the SOHO spacecraft \citep{UGR2000}.
%
%
We sampled these data with the same window function and weights as the
\acena{} observations, and extracted the frequencies of the strongest 30
oscillation peaks using the same iterative sine-wave fitting method as for
\acena{} (see Section~\ref{sec.freq}).  We compared these frequencies with
the values measured by \citet{TBGB2000} from the full 805-day GOLF series,
and thereby determined the scatter on our ``observed'' solar frequencies.
We then used the calibration derived from the simulations, as described in
the previous paragraph, to infer mode lifetimes.  The results for two
frequency ranges are given in Table~\ref{tab.lifetime}.  In this case, we
were able to compare these mode lifetimes with published values.  For this
we took values measured by \citet{CEI97} from fitting Lorentzian profiles
to a 32-month power spectrum of ground-based solar velocities.  We see that
our method reproduces the published lifetimes within the uncertainties,
confirming the validity of the technique.

Our main result is that mode lifetimes for \acena{} are 1--2\,d,
significantly shorter than the values of 3--4\,d found in the Sun.  Is this
result consistent with the results for \acena{} presented by
\citet{B+C2002}?  Their frequencies were measured from a time series of
considerably longer duration than ours (12.4\,d versus 4.6\,d), so we would
expect them to have lower scatter.  This is indeed the case -- we measure a
scatter of about 0.8\,\muHz{} from their frequencies.  This is much greater
than would be expected for coherent oscillations (about 0.1--0.2\,\muHz),
but is consistent with the mode lifetime that we measured from our
observations.


Note that the mode lifetime~$\tau$ is related to the FWHM~$\Gamma$ of the
corresponding Lorentzian via $\tau = 1/(\pi\Gamma)$.  Thus, if \acena{}
were observed for long enough, we would expect each mode to generate a
Lorentzian profile in the power spectrum with width $\Gamma = 2$--3\,\muHz.
This is less than the FWHM of 4\,\muHz{} of the spectral window of our
observations, which explains why such a profile is not apparent in our
power spectrum.

 We have assumed that the scatter of the observed frequencies about a
smooth trend is due to the effects of the finite mode lifetime.  However,
we should also consider that the actual eigenfrequencies of the star may
have small deviations from regularity.  For the Sun, this is indeed the
case, but at a low level: the frequencies reported by \citet{TBGB2000}
deviate from a parabola in the echelle diagram by 0.2--0.3\,\muHz.  If
\acena{} possesses irregularities of this magnitude, their contribution to
the scatter that we measured is very small.  Correcting for this would lead
us to reduce the values of scatter in Table~\ref{tab.lifetime} by about
0.02\,\muHz, and to increase the estimates of mode lifetimes by a few
percent.  We conclude that such frequency deviations have negligible effect
on our measurements of the mode lifetime.

\section{Conclusions}

We have analysed the two-site observations of \acena{} that were described
in Paper~I.  Our main results are as follows:
\begin{enumerate}

\item By adjusting the weights for each night, we optimized the window
function and reduced the sidelobes in the spectral window to only 3.6\% in
power.  This allowed us to measure the frequencies of the strongest peaks
in the power spectrum using iterative sine-wave fitting, without any
ambiguities from daily aliases.

\item We clearly detected several modes with $l=3$, the first time this has
been achieved for solar-like oscillations.  

\item We made fits to the measured frequencies and obtained the relations
in equations~(\ref{eq.nu0})--(\ref{eq.nu3}).  Given the significant scatter
in the frequencies, we suggest that comparison with theoretical models is
best done via these fitted relations.

\item Based on the scatter of the observed frequencies, we inferred mode
lifetimes in \acena{} of 1--2\,d.  These shorter-than-expected lifetimes
present a challenge to theoretical models \citep{HBChD99}, and also raise
concerns for asteroseismology because they degrade the precision with which
oscillation frequencies can be measured.  It is clear that more stars need
to be observed before we can determine how this important parameter varies
in the H-R diagram.

\end{enumerate}

\acknowledgments

This work was supported financially by the Australian Research Council, by
the Danish Natural Science Research Council and by the Danish National
Research Foundation through its establishment of the Theoretical
Astrophysics Center.  We further acknowledge support by NSF grant
AST-9988087 (RPB), and by SUN Microsystems.  We thank Fran\c{c}ois Bouchy
and Ian Roxburgh for valuable comments on this paper.

\clearpage

\clearpage

\begin{table*}
\small
\caption{\label{tab.freqs} Oscillation frequencies for \acena{} (\muHz)}
\begin{center}
\begin{tabular}{ccccc}
\noalign{\medskip}
\tableline
\tableline
\noalign{\smallskip}
~~~$n$~~~ &  ~~~$l=0$~~~  & ~~~$l=1$~~~  & ~~~$l=2$~~~  & ~~~$l=3$~~~ \\
\tableline
\noalign{\smallskip}
 14 &  \ldots & 1675.9 & \ldots & \ldots \\
 15 &  \ldots & 1779.7 & 1828.6 & \ldots \\
 16 &  1839.2 & 1885.9 & 1935.7 & \ldots \\
 17 &  1943.3 & 1993.8 & 2038.9 & 2082.9 \\
 18 &  2045.5 & 2094.6 & 2146.3 & 2193.1 \\
 19 &  2152.9 & 2203.2 & 2253.4 & 2296.3 \\
 20 &  2258.1 & 2309.1 & 2357.3 & 2404.8 \\
 21 &  2364.0 & 2412.4 & 2463.4 & 2507.5 \\
 22 &  2471.5 & 2522.1 & 2572.7 & 2616.8 \\
 23 &  2572.7 & 2627.1 & 2676.8 & 2723.5 \\
 24 &  2682.7 & 2733.2 & 2783.4 & \ldots \\
 25 &  \ldots & 2840.2 & \ldots & \ldots \\
 26 &  2895.9 & 2945.7 & 2998.3 & \ldots \\
 27 &  \ldots & 3055.1 & \ldots & \ldots \\
\noalign{\smallskip}
\tableline
\end{tabular}
\end{center}
\end{table*}

\begin{table*}
\small
\caption{\label{tab.amps} Oscillation amplitudes for \acena{} (\cms)}
\begin{center}
\begin{tabular}{ccccc}
\noalign{\medskip}
\tableline
\tableline
\noalign{\smallskip}
~~~$n$~~~ &  ~~~$l=0$~~~  & ~~~$l=1$~~~  & ~~~$l=2$~~~  & ~~~$l=3$~~~ \\
\tableline
\noalign{\smallskip}
 14 &  \ldots & ~~6.5  & \ldots & \ldots \\
 15 &  \ldots &  11.8  & ~~8.8  & \ldots \\
 16 & 13.5    &  25.0  &  30.2  & \ldots \\
 17 & 14.4    & ~~5.9  &  19.1  &  10.6  \\
 18 & 13.3    &  24.9  &  18.7  &  24.8  \\
 19 & 34.0    &  18.6  &  21.3  &  15.2  \\
 20 & 20.7    &  31.3  &  31.8  &  14.4  \\
 21 & 21.1    &  31.6  &  27.1  &  19.6  \\
 22 & 16.7    &  18.5  &  40.3  &  18.9  \\
 23 & 40.3    &  26.2  &  13.9  &  11.2  \\
 24 & 27.8    &  30.5  &  18.0  & \ldots \\
 25 &  \ldots &  14.4  & \ldots & \ldots \\
 26 & 21.4    &  14.4  &  16.8  & \ldots \\
 27 &  \ldots & ~~7.5  & \ldots & \ldots \\
\noalign{\smallskip}
\tableline
\end{tabular}
\end{center}
\end{table*}

\begin{table*}
\small
\caption{\label{tab.lifetime} Mode lifetimes in \acena{} and the Sun}
\begin{center}
\begin{tabular}{lcccc}
\noalign{\medskip}
\tableline
\tableline
\noalign{\smallskip}
  &    Frequency   &   Scatter    &    Inferred & Published \\
  &     (\muHz)  &    (\muHz)   &     lifetime (d) & lifetime (d) \\
\tableline
\noalign{\smallskip}
 \acena &  1700--2400   &   1.50 $\pm$    0.23 & 
    1.4$^{\scriptscriptstyle{+0.5}}_{\scriptscriptstyle{-0.4}}$ \\
\noalign{\smallskip}
        &  2400--3000   &   1.60 $\pm$    0.25 & 
    1.3$^{\scriptscriptstyle{+0.5}}_{\scriptscriptstyle{-0.4}}$ \\
\noalign{\smallskip}
\tableline
\noalign{\smallskip}
 Sun    &  2400--3050   &   0.95 $\pm$    0.17 & 
    4.0$^{\scriptscriptstyle{+2.0}}_{\scriptscriptstyle{-1.1}}$ &
    3.17 $\pm$ 0.15 \\
\noalign{\smallskip}
        &  3050--3700   &   1.31 $\pm$    0.23 & 
    2.0$^{\scriptscriptstyle{+1.0}}_{\scriptscriptstyle{-0.6}}$ &
    1.59 $\pm$ 0.09 \\
\tableline
\end{tabular}
\end{center}
\end{table*}

\end{document}